\documentclass[pra,aps,showpacs,superscriptaddress,twocolumn]{revtex4}

\usepackage{amsmath}
\usepackage{bm}
\usepackage{graphicx}

\begin{document}

\title{Controlled generation and manipulation of vortex dipoles in a
	Bose-Einstein condensate}

\author{Tomohiko Aioi}
\affiliation{Department of Engineering Science, University of
Electro-Communications, Tokyo 182-8585, Japan}

\author{Tsuyoshi Kadokura}
\affiliation{Department of Engineering Science, University of
Electro-Communications, Tokyo 182-8585, Japan}

\author{Tetsuo Kishimoto}
\affiliation{Department of Engineering Science, University of
Electro-Communications, Tokyo 182-8585, Japan}
\affiliation{Center for Frontier Science and Engineering, University of
Electro-Communications, Tokyo 182-8585, Japan}
\affiliation{PRESTO, Japan Science an Technology Agency (JST), Saitama
	332-0012, Japan}

\author{Hiroki Saito}
\affiliation{Department of Engineering Science, University of
Electro-Communications, Tokyo 182-8585, Japan}

\date{\today}

\begin{abstract}
We propose methods to generate and manipulate vortex dipoles in a
Bose-Einstein condensate using Gaussian beams of red or blue-detuned
laser.
Velocity-controlled vortex dipoles are shown to be created and launched by
a red-detuned beam and by two blue-detuned beams.
Critical velocities for the vortex nucleation are investigated.
The launched vortex dipoles can be trapped, curved, accelerated, and
decelerated using Gaussian beams.
Collisions between vortex dipoles are demonstrated.
\end{abstract}

\pacs{03.75.Lm, 03.75.Kk, 67.85.De, 47.37.+q}

\maketitle

\section{Introduction}

A vortex dipole is a pair of vortices with opposite circulations, which is
a fundamental excitation in a two dimensional (2D) fluid and plays a
pivotal role in thermodynamic behavior of 2D
systems~\cite{Berezinskii,Kosterlitz}.
In superfluids, vortices are topological defects of the macroscopic wave
function and a vortex dipole consists of a quantized vortex and
antivortex, whose circulations are quantized to $\pm h / m$ with $h$ and
$m$ being Planck's constant and the mass of a
particle~\cite{Onsager,Feynman}.
Such a topological structure of a quantized vortex dipole has been
observed in a Bose-Einstein condensate (BEC) of an atomic
gas~\cite{Inouye} and in an exciton-polariton condensate~\cite{Roumpos}.
Recently, dynamics of vortex dipoles have directly been observed in oblate
BECs~\cite{Neely,Freilich,Middel}.

In the experiments reported in Refs~\cite{Inouye,Neely}, vortex dipoles
were created by stirring a BEC with a blue-detuned laser beam.
When the velocity of the beam exceeds a critical velocity,
vortex-antivortex pairs are released from the density dip produced by the
laser beam.
The created vortex dipoles then move through the BEC and exhibit various
interesting dynamics~\cite{Neely,Freilich,Middel,Kuopan}, which can be
observed in real time using the technique developed by the Amherst
group~\cite{Freilich,Middel}.
Theoretical study on a moving potential in superfluids has so far been
focused on drag force~\cite{Frisch,Wini99,Wini00,Aftalion},
vortex formation and shedding dynamics~\cite{Jackson,Nore,Sasaki},
critical velocity~\cite{Crescimanno,Sties}, 
scaling laws~\cite{Huepe}, and
potential oscillation~\cite{Fujimoto}.
In addition to these studies, it is important to establish methods for
controlled generation and manipulation of vortex dipoles, which increases
the range of experiments and leads to precise understanding of dynamical
properties of vortex dipoles.

The aim of the present paper is to propose methods to generate and
manipulate quantized vortex dipoles in a BEC using not only a single
blue-detuned laser beam but also a red-detuned beam and multiple beams.
We will show that dynamics of vortex dipole creation for a red-detuned
beam (attractive potential for atoms) are quite different from those for a
blue-detuned beam (repulsive potential for atoms).
We propose methods to create and launch vortex dipoles with a controlled
velocity and angle using a red-detuned beam or two blue-detuned beams.
We also show that two red-detuned beams can trap a vortex dipole between
them.
The trajectories of the launched vortex dipoles can be curved by both
blue- and red-detuned beams.
The launched vortex dipoles can also be accelerated and decelerated by
Gaussian beams with time-dependent intensity.
Using two vortex-dipole launchers, we demonstrate collisions of vortex
dipoles.

The present paper is organized as follows.
Section~\ref{s:form} provides the formulation and numerical method.
Section~\ref{s:red} studies vortex dipole generation by a red-detuned beam
and shows that it can launch a vortex dipole in a controlled manner.
Section~\ref{s:double} investigates dynamics for two laser beams.
Section~\ref{s:manip} proposes methods to manipulate launched vortex
dipoles and demonstrate their collisions.
Section~\ref{s:conc} gives conclusions to this study.

\section{Formulation of the problem}
\label{s:form}

We consider a zero-temperature BEC in the mean-field approximation.
The dynamics of the macroscopic wave function $\Psi$ is described by the
Gross-Pitaevskii (GP) equation given by
\begin{equation} \label{GP}
i \hbar \frac{\partial \Psi}{\partial t} = -\frac{\hbar^2}{2m} \nabla^2
\Psi + [V(x, y, t) + V_z(z)] \Psi + \frac{4 \pi \hbar^2 a}{m} |\Psi|^2
\Psi,
\end{equation}
where $m$ is the atomic mass, $V$ and $V_z$ are the external potentials,
and $a$ is the $s$-wave scattering length.
We assume that the system is tightly confined in the $z$ direction and the
confinement energy is much larger than other energy scales of the system.
The wave function is thus frozen into $\Psi(x, y, z, t) = \psi(x, y, t)
e^{i E_0 t / \hbar} \psi_0(z)$, where $\psi_0(z)$ and $E_0$ are the ground
state and its energy for $V_z(z)$.
Multiplying $\psi_0(z)$ to Eq.~(\ref{GP}) and integrating it with respect
to $z$, we reduce the 3D GP equation to 2D as
\begin{equation} \label{GP2D}
i \hbar \frac{\partial \psi}{\partial t} = -\frac{\hbar^2}{2m}
\nabla_\perp^2 \psi + V(x, y, t) \psi + g |\psi|^2 \psi,
\end{equation}
where $\nabla_\perp^2$ is the 2D Laplacian and $g = 4 \pi \hbar^2 a m^{-1}
\int |\psi_0|^4 dz$ is the effective interaction parameter in 2D space.
We normalize the 2D wave function as $\tilde \psi = n_0^{-1/2} \psi$,
where $n_0$ is the 2D atomic density in the absence of $V$.
Normalizing the length and time by $\xi = \hbar / (m g n_0)^{1/2}$ and
$\tau = \xi / v_{\rm s} = \hbar / (g n_0)$, where $v_{\rm s} = (g n_0 /
m)^{1/2}$ is the sound velocity, Eq.~(\ref{GP2D}) is rewritten as
\begin{equation} \label{GP2Dn}
i \frac{\partial \tilde\psi}{\partial \tilde t} = -\frac{1}{2}
\tilde\nabla_\perp^2 \tilde\psi + \tilde V \tilde \psi + |\tilde\psi|^2
\tilde\psi,
\end{equation}
where the tilde represents normalized quantity.
From this normalization, we find that the scaling law holds for the
mean-field approximation.
For example, the dynamics for the interaction strength $g n_0$ with a
Gaussian potential $V = V_0 \exp\{ -[(x - v t)^2 + y^2] / d^2 \}$ and that 
for $\alpha g n_0$ with $V = \alpha V_0 \exp\{ -[(x - \sqrt{\alpha} v t)^2
+ y^2] / (d / \sqrt{\alpha})^2 \}$ obey the same equation (\ref{GP2Dn})
for an arbitrary constant $\alpha > 0$.

In the numerical simulations, we first prepare the ground state for $V(x,
y, t = 0)$ by the imaginary time propagation of Eq.~(\ref{GP2Dn}), i.e.,
$i$ is replaced by $-1$ on the left-hand side of Eq.~(\ref{GP2Dn}).
We then add a small white noise to the ground state to break the exact
numerical symmetry.
Starting from this initial state, we obtain the time evolution by solving
Eq.~(\ref{GP2Dn}).
The imaginary and real time propagation is calculated by the
pseudospectral method, and therefore the periodic boundary condition is
imposed.
We take a large enough space in the simulations so that the boundary does
not affect the dynamics.

\section{Vortex generation by a red-detuned Gaussian beam}
\label{s:red}

\subsection{Flow around a moving potential}

First, we consider a single Gaussian laser beam moving through the
condensate, which produces a potential as
\begin{equation} \label{gaussian}
V(x, y, t) = V_0 \exp \left[-\frac{(x + v t)^2 + y^2}{d^2} \right],
\end{equation}
where $d$ is the $1/e$ width of the laser beam.
We assume that the laser beam moves in the $-x$ direction with the
velocity $v$.
The constant $V_0$ is proportional to the intensity of the laser beam,
which is positive for blue detuning and negative for red detuning with
respect to the resonant frequency of the electric dipole transition of
atoms.
The detuning is large and the heating of atoms by the spontaneous emission
is negligible.

Frisch {\it et al.}~\cite{Frisch} showed that quantized vortex dipoles are
created around a moving rigid disk ($V = +\infty$ inside a disk) above a
critical velocity.
The vortex and antivortex are created at the lateral sides of the disk,
and they are released into the wake (see Fig.~1 of Ref.~\cite{Frisch}).
The Gaussian potential (\ref{gaussian}) for $V_0 > 0$ also exhibits
similar behavior~\cite{Jackson,Wini00}.

\begin{figure}[t]
\includegraphics[width=8.5cm]{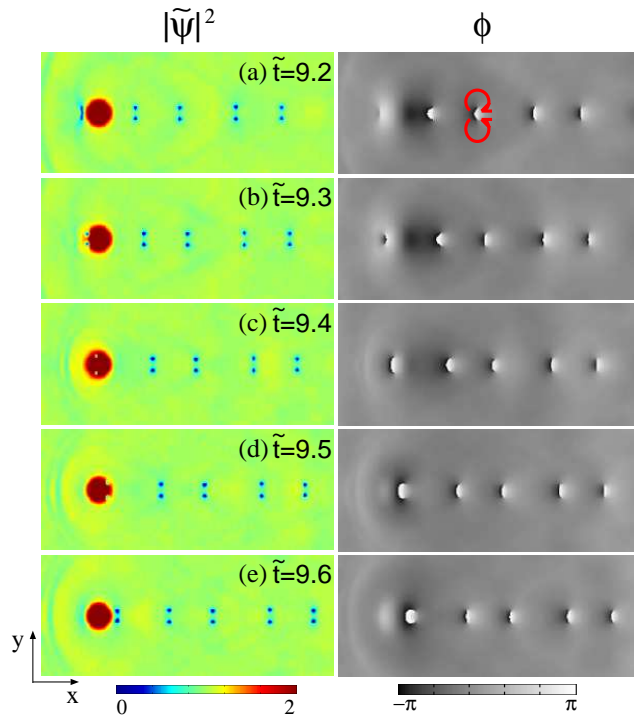}
\caption{
Dynamics of the density $|\tilde\psi|^2 = |\psi|^2 / n_0$ and phase $\phi
= {\rm arg} \psi$ profiles for the Gaussian potential with $V_0 / (g n_0)
= -5$, $v / v_{\rm s} = 0.6$, and $d / \xi = 4.5$ in the frame comoving
with the potential.
The arrows indicate the directions of circulations.
The field of view of each panel is $160 \xi \times 67 \xi$.
}
\label{f:single}
\end{figure}
By contrast, we investigate the dynamics for a red-detuned $(V_0 < 0)$
moving Gaussian beam, which is shown in Fig.~\ref{f:single}.
First, a density dip appears at the head of the potential
[Fig.~\ref{f:single} (a)], which splits into vortex and antivortex
[Fig.~\ref{f:single} (b)].
The vortex pair then passes through the potential avoiding the high
density region [Fig.~\ref{f:single} (c)], and it is released behind the
potential.
The released vortex dipole moves in the $-x$ direction with the velocity
$\hbar / (m l) = v_s \xi / l$, where $l$ is the distance between the
vortex and antivortex.
Such vortex dynamics are repeated and vortex dipoles are shed in the wake
periodically.
The train of vortex dipoles is dynamically unstable and eventually sinuous
destabilization spreads it in the $\pm y$
directions~\cite{Nore93,Sasaki}.
In contrast to the case of a blue-detuned Gaussian beam, the B\'enard-von
K\'arm\'an vortex street~\cite{Sasaki} was not found for a red-detuned
beam to the best of our investigation.

For an attractive potential ($V_0 < 0$), a vortex dipole is nucleated at
the head of a moving potential, whereas a vortex dipole is nucleated at
the lateral sides for a repulsive potential ($V_0 > 0$).
To understand these behaviors, we consider the irrotational flow of an
inviscid and incompressible fluid around a rigid disk.
The velocity field in the frame comoving with the disk is given
by~\cite{Lamb}
\begin{equation}
v_x - i v_y = U \left( 1 - \frac{R^2}{r^2} e^{-2i\theta} \right),
\end{equation}
where $U$ and $R$ are the velocity and radius of the disk, $r = (x^2 +
y^2)^{1/2}$, and $\theta = {\rm arg} (x + i y)$.
The velocity becomes maximum at $(r, \theta) = (R, \pm \pi / 2)$, which
agree with the locations at which vortices are nucleated.
This is because vortices are nucleated when the velocity exceeds the local
Landau critical velocity.
Moreover, the density is low where the velocity is large due to
Bernoulli's law and a quantized vortex with a density hole is easy to be
nucleated.

For a disk-shaped strong attractive potential, where the density inside
the disk $n_{\rm in}$ is much larger than the outside density $n_{\rm
out}$, the inside velocity must vanish in the limit of $n_{\rm in} /
n_{\rm out} \rightarrow \infty$ since the normal component of the flow
must be continuous across the boundary, $n_{\rm in} v_{\rm in}^\perp =
n_{\rm out} v_{\rm out}^\perp$.
Therefore, the tangential component of the velocity $v^\parallel$ must
vanish at the boundary, since it must be continuous.
Such velocity field outside the disk is given by
\begin{equation} \label{negpot}
v_x - i v_y = U \left( 1 + \frac{R^2}{r^2} e^{-2i\theta} \right).
\end{equation}
The velocity becomes maximum at $r = R$ and $\theta = 0, \pi$, and thus a
vortex dipole is nucleated at the head of a potential.
Perturbations at the tail of a potential ($r = R$ and $\theta = 0$) do not
grow because they go downstream away.

\begin{figure}[t]
\includegraphics[width=8.5cm]{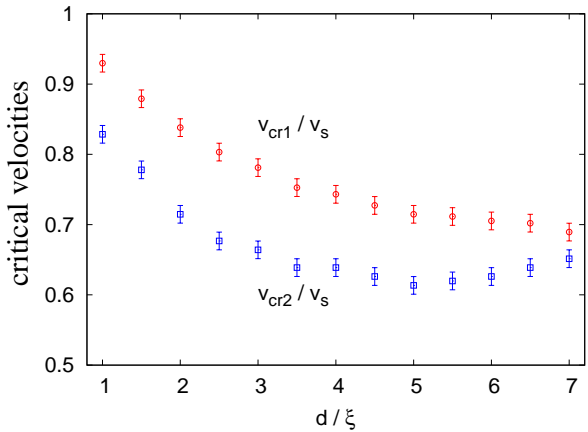}
\caption{
Critical velocities for vortex dipole creation versus beam waist $d$ of a
moving red-detuned Gaussian beam.
When the velocity of the beam is adiabatically increased, the vortex
dipole creation occurs at $v_{{\rm cr} 1}$ (circles).
When the velocity of the beam is adiabatically decreased, the vortex
dipole creation stops at $v_{{\rm cr} 2}$ (squares).
The intensity of the Gaussian potential (\ref{gaussian}) is fixed to $V_0
/ (g n_0) = -5$.
}
\label{f:phase}
\end{figure}
It is interesting to note that the system exhibits hysteresis behavior
with respect to the velocity.
When the velocity of the beam is adiabatically increased from zero, the
successive vortex dipole creation starts at a critical velocity $v_{{\rm
cr} 1}$, which is shown by the circles in Fig.~\ref{f:phase}.
On the other hand, when the velocity of the beam is decreased from $v >
v_{{\rm cr} 1}$, the vortex dipole generation does not stop at $v =
v_{{\rm cr} 1}$.
The velocity $v_{{\rm cr} 2}$ (squares in Fig.~\ref{f:phase}) at which the
vortex dipole generation stops is slower than $v_{{\rm cr} 1}$.
For the velocity $v_{{\rm cr} 2} < v < v_{{\rm cr} 1}$, the stationary
flow remains to be stationary, and the successive vortex dipole creation
is kept once it is triggered by perturbations.

\subsection{Launching a vortex dipole}
\label{s:launch}

\begin{figure}[t]
\includegraphics[width=8.5cm]{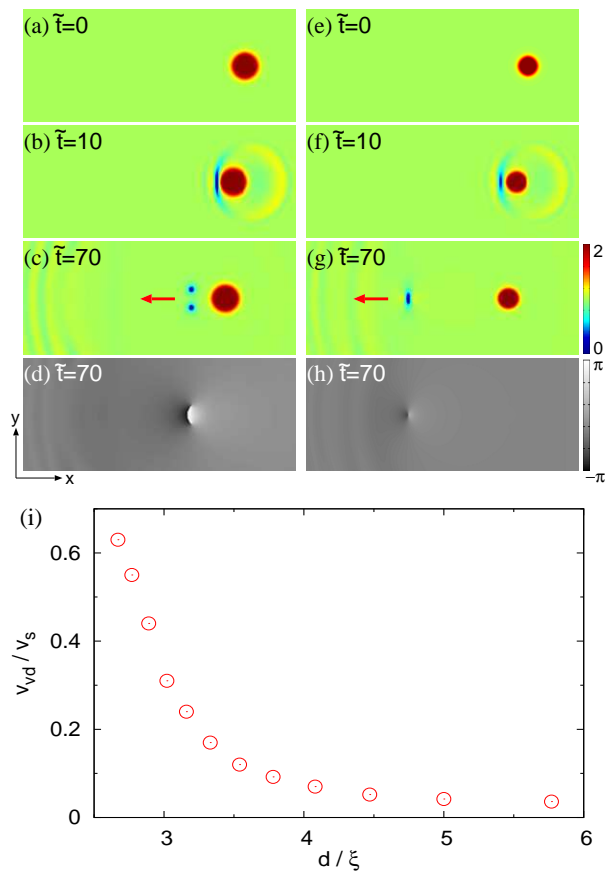}
\caption{
(a)-(h) Snapshots of the density $|\psi|^2 / n_0$ and phase $\phi = {\rm
arg} \psi$ profiles in the rest frame.
A Gaussian potential with $V_0 / (g n_0) = -5$ is moved as in
Eq.~(\ref{displace}) with $v_0 / v_{\rm s} = 0.47$ and $T / \tau = 17$.
The beam waist is $d / \xi = 3.54$ in (a)-(d) and $d = 2.67$ in (e)-(h).
The arrows in (c) and (g) indicate the traveling direction of the vortex
dipoles.
The field of view of each panel is $78 \xi \times 33 \xi$.
(i) Velocity of the launched vortex dipole versus beam waist $d$.
The parameters except $d$ are the same as those in (a)-(h).
}
\label{f:launch}
\end{figure}
Using a red-detuned Gaussian laser beam, we can launch vortex dipoles in
a controlled manner.
The initial state is the ground state for the Gaussian potential in
Eq.~(\ref{gaussian}) with $V_0 < 0$ and $v = 0$.
The potential then starts to move in the $-x$ direction at $t = 0$ and
stops at $t = T$ as
\begin{equation} \label{displace}
v = \left\{ \begin{array}{ll} 0 & (t < 0, t > T) \\
v_0 & (0 \leq t \leq T) \end{array} \right..
\end{equation}
Figures~\ref{f:launch} (a)-\ref{f:launch} (h) show launching dynamics of a
vortex dipole in the rest frame.
After the stop of the potential, the created vortex dipole is released
from the potential and continues to move in the $-x$ direction.
Figures~\ref{f:launch} (e)-\ref{f:launch} (h) show the case of a smaller
potential, in which a vortex dipole is bound so tight that the topological
defects cannot be discerned [Fig.~\ref{f:launch} (h)].
Such a solitonic structure is referred to as a rarefaction
pulse~\cite{Jones}, which has been shown to be formed by a local
depletion~\cite{Berloff} and an oscillating potential~\cite{Fujimoto}.
The rarefaction pulse propagates faster than the vortex dipole
[Figs.~\ref{f:launch} (c) and \ref{f:launch} (g)].
Figure~\ref{f:launch} (i) shows the dependence of the velocity $v_{\rm
vd}$ of the launched vortex dipole on the beam waist $d$.
Since the distance between the created vortex and antivortex increases
with an increase in the size of the potential, the velocity of the vortex
dipole is a decreasing function of the size of the potential.
Thus, the vortex dipole can be launched with controlled velocity.

\section{Vortex generation by multiple Gaussian beams}
\label{s:double}

\subsection{Flow around double beams}
\label{s:double1}

We next consider the case of multiple Gaussian potentials.
For simplicity, we assume that two identical Gaussian potentials move at
the same velocity as
\begin{equation} \label{twogauss}
V = V_0 \sum_{j=1}^2 \exp
\left[-\frac{(x - x_j + v t)^2 + (y - y_j)^2}{d^2} \right].
\end{equation}
The positions of the potentials are, without loss of generality,
$(x_j, y_j) = (\mp \frac{D}{2} \sin \chi, \pm \frac{D}{2} \cos \chi)$ with
the upper signs for $j = 1$ and the lower signs for $j = 2$, where the
angle $\chi$ determines the alignment of the potentials.

\begin{figure}[t]
\includegraphics[width=8.5cm]{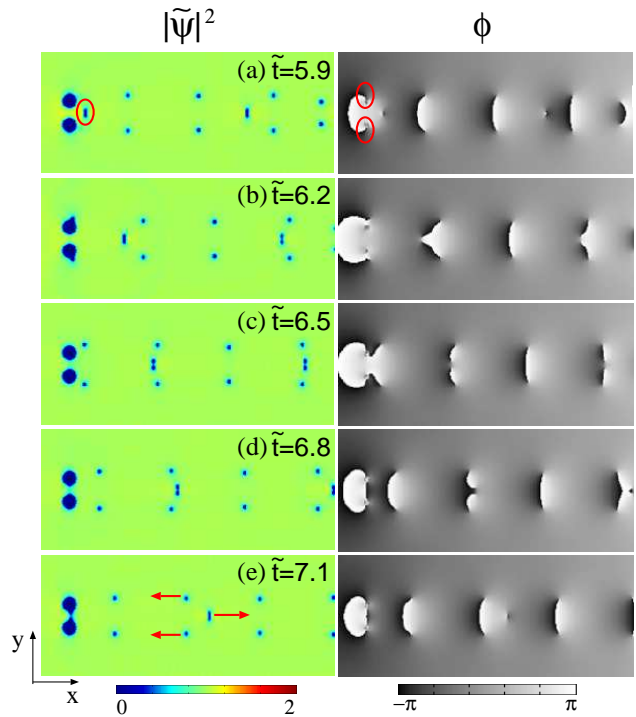}
\caption{
Dynamics of the density $|\tilde\psi|^2 = |\psi|^2 / n_0$ and phase $\phi
= {\rm arg} \psi$ profiles for two Gaussian potentials with $V_0 / (g
n_0) = 100$, $d / \xi = 1.58$, $v / v_{\rm s} = 0.35$, and $D / \xi =
12.6$ in the frame comoving with the potential.
The circle in the left panel of (a) indicates a vortex dipole released
from between the potentials, and those in the right panel its counterpart
remained in the potentials.
The arrows in (e) indicate the directions in which the vortices
propagate.
The field of view of each panel is $160 \xi \times 67 \xi$.
}
\label{f:double}
\end{figure}
Figure~\ref{f:double} shows a typical dynamics for two blue-detuned ($V_0
> 0$) beams with $\chi = 0$.
The initial state is the ground state for $v = 0$ and the potentials start
to move in the $-x$ direction at $t = 0$.
After the initial disturbance, the system exhibits periodic behavior.
A vortex dipole is created between the potentials and released in the wake
[circle in the left panel of Fig.~\ref{f:double} (a)] with its counter
part being remained in the potentials [circles in the right panel of
Fig.~\ref{f:double} (a)].
Then, the topological defects remained in the potentials are released from
both ends [Fig.~\ref{f:double} (c)], and this dynamics is repeated.
We note that small vortex dipoles released from between the potentials
propagate in the $x$ direction, whereas vortices in the outer rows
propagate in the $-x$ direction [arrows in Fig.~\ref{f:double} (e)].

\begin{figure}[t]
\includegraphics[width=8.5cm]{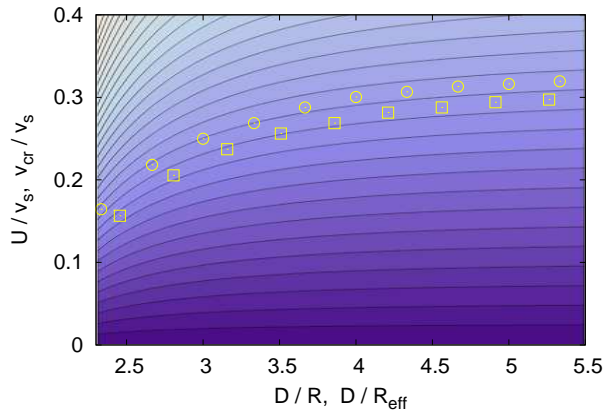}
\caption{
Contour lines show the maximum velocity $v_{\rm max}$ as a function of $D
/ R$ and $U / v_{\rm s}$, obtained from Eq.~(\ref{vf}).
The contour lines indicate $v_{\rm max} / v_{\rm s} = 0.05, 0.1, 0.15,
0.2, \cdots$ from below.
The plots show the critical velocity $v_{\rm cr}$ for
vortex dipole nucleation versus $D / R_{\rm eff}$, which is obtained by
simulations as in Fig.~\ref{f:double}.
Here $R_{\rm eff}$ is the radius of the density hole produced by the
potential.
The beam waist and the effective radius are $(d / \xi, R_{\rm eff} / \xi)
= (5, 9.5)$ for the circles and $(10, 18)$ for the squares.
}
\label{f:critical}
\end{figure}

In order to understand the behavior in Fig.~\ref{f:double}, we consider
the irrotational flow of an inviscid and incompressible fluid around two
rigid disks with radius $R > D / 2$ located at $(x, y) = (\mp \frac{D}{2}
\sin \chi, \pm \frac{D}{2} \cos \chi)$.
The velocity field is given by~\cite{Hicks,Greenhill,Crowdy}
\begin{eqnarray} \label{vf}
v_x - i v_y & = & \frac{4 U \sinh^2 \beta}{(z / R)^2 + \sinh^2 \beta}
\biggl[ e^{i \chi} \wp(i \tau - i \beta) 
\nonumber \\
& & + e^{-i \chi} \wp(i \tau + i \beta)
+ \frac{2}{\pi} \zeta(\pi) \cos \chi \biggr],
\end{eqnarray}
where $\beta = \cosh^{-1}[D / (2 R)]$, $z = x + i y$, $\tau = \beta +
\log(i e^{-i \chi} z / R + \sinh \beta) - \log(i e^{-i \chi} z / R - \sinh
\beta)$, and $\wp$ and $\zeta$ are Weierstrass's elliptic functions with
half periods $\pi$ and $i \cosh^{-1} [(D^2 / R^2 - 2) / 2]$.
The velocity field (\ref{vf}) approaches to $v_x - i v_y = U$ for $|z|
\rightarrow \infty$.
For $\chi = 0$, which corresponds to the case in Fig.~\ref{f:double}, the
velocity field in Eq.~(\ref{vf}) is maximum at the inner edges of the
disks.
We denote the maximum velocity as $v_{\rm max} \equiv v_x(x = 0, y = \pm D
/ 2 \mp R)$.
The contour plot in Fig.~\ref{f:critical} shows $v_{\rm max}$ as a
function of the distance $D$ between the disks and the unperturbed
velocity $U$.
The maximum velocity $v_{\rm max}$ monotonically decreases with an
increase in $D$ and linearly increases with an increase in $U$.

In Fig.~\ref{f:critical}, we superimpose the plots of the critical
velocity $v_{\rm cr}$ for vortex dipole creation, obtained from the GP
equation (\ref{GP2Dn}) numerically.
In the time evolution, we adiabatically increase the velocity $v$ of the
potential from $v = 0$ until a vortex dipole is nucleated as in
Fig.~\ref{f:double}, which determines the critical velocity $v_{\rm cr}$. 
From Fig.~\ref{f:critical}, we find that the plots roughly fit the contour
lines, implying that a vortex dipole is nucleated when the maximum
velocity between the disks exceeds a local critical velocity, which is
almost independent of $D$.

\subsection{Launching a vortex dipole}

\begin{figure}[t]
\includegraphics[width=8.0cm]{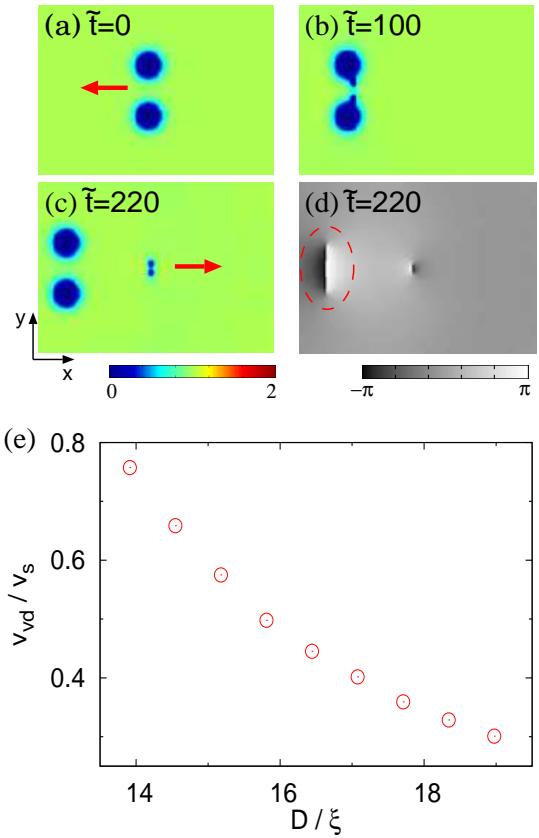}
\caption{
(a)-(c) Dynamics of the density profiles $|\psi|^2 / n_0$ for two Gaussian
potentials with $V_0 / (g n_0) = 5$, $d / \xi = 4.5$, and $D / \xi = 19$
in the rest frame.
The potentials are moved as in Eq.~(\ref{displace}) with $v_0 / v_{\rm s}
= 0.24$ and $T / \tau = 133$.
The arrows in (a) and (c) indicate the direction of the displacement of
the potentials and that of the propagation of the vortex dipole,
respectively.
(d) Phase profile, ${\rm arg} \psi$, at $\tilde t = t / \tau = 220$.
The topological defects in the dashed circle are the counterpart of the
launched vortex dipole.
The field of view of each panel is $90 \xi \times 65 \xi$.
(e) Velocity of the launched vortex dipole versus distance $D$ between the
two Gaussian potentials.
The parameters except $D$ are the same as those in (a)-(d).
}
\label{f:launch2}
\end{figure}
Using two blue-detuned Gaussian beams, we can launch a vortex dipole, as
in the case of a red-detuned beam in Sec.~\ref{s:launch}.
The initial state is the ground state in the presence of two Gaussian
potentials (\ref{twogauss}) with $V_0 > 0$, $v = 0$, and $\chi = 0$.
The potentials are then displaced as in Eq.~(\ref{displace}).
Figures~\ref{f:launch2} (a)-\ref{f:launch2} (c) show the launching
dynamics in the rest frame.
A vortex dipole is generated between the potential and moves in the
direction opposite to that of the displacement of the potentials.
A vortex dipole is thus launched by ``pulling'' the potential, which is in
contrast to the case of a red-detuned beam, in which a vortex dipole is
launched by ``pushing'' the potential.
After the launch, a counterpart of the released vortex-antivortex pair
remains at the potential  [dashed circle in Fig.~\ref{f:launch2} (d)],
which prevents the successive launch.
By contrast, the successive launch is possible for a red-detuned beam.

Figure~\ref{f:launch2} (e) shows the dependence of the velocity $v_{\rm
vd}$ of a launched vortex dipole on the distance $D$ between the two
Gaussian potentials.
The velocity $v_{\rm vd}$ decreases with an increase in $D$, since the
distance between the launched vortex and antivortex increases.

\subsection{Trapping of a vortex dipole}

\begin{figure}[t]
\includegraphics[width=8.5cm]{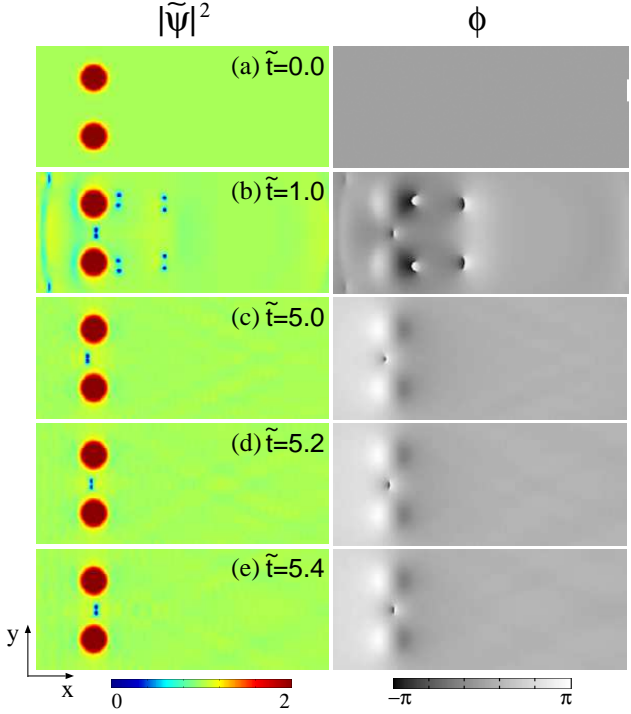}
\caption{
Dynamics of the density $|\tilde\psi|^2 = |\psi|^2 / n_0$ and phase $\phi
= {\rm arg} \psi$ profiles for the potential in Eq.~(\ref{twogauss}) with
$V_0 / (g n_0) = -5$, $v / v_{\rm s} = 0.6$, $d / \xi = 4.5$, and $D / \xi
= 31.6$ in the frame comoving with the potential.
The field of view of each panel is $160 \xi \times 67 \xi$.
}
\label{f:trap}
\end{figure}
We consider the dynamics for two red-detuned beams as shown in
Fig~\ref{f:trap}.
The initial state is the ground state for two Gaussian potentials in
Eq.~(\ref{twogauss}) with $V_0 < 0$, $v = 0$, and $\chi = 0$
[Fig~\ref{f:trap} (a)] and the potentials start to move in the $-x$
direction at $t = 0$.
Initially, the sudden increase in the velocity disturbs the system and
several vortex dipoles are generated near the potential [Fig~\ref{f:trap}
(b)].
The vortex generation is only at the starting, and the generated vortex
dipoles depart from the potential except the one between the potentials.
Interestingly, this vortex dipole is trapped between the potentials and
oscillates around the center in the moving frame [Figs~\ref{f:trap}
(c)-\ref{f:trap} (e)].

In order to understand the trapping of a vortex dipole between moving
potentials as shown in Fig.~\ref{f:trap}, we perform linear stability
analysis assuming inviscid, incompressible, and irrotational flow.
We consider the problem in the frame comoving with the potentials.
Two point vortices with circulations $-h / m \equiv -\kappa$ and
$\kappa$ are located at $(x_1, y_1) = (\xi_1, l / 2 + \eta_1)$ and $(x_2,
y_2) = (\xi_2, -l / 2 + \eta_2)$, where $\xi_1$, $\eta_1$, $\xi_2$, and
$\eta_2$ are small deviations from the stationary state and $l$ is the
distance between the vortex and antivortex for the stationary state.
Equations of motion for the deviations are given by $(j = 1, 2)$
\begin{subequations} \label{eom}
\begin{eqnarray}
\dot{\xi}_j & = & v_x(x_j, y_j) - \frac{\kappa}{2\pi} \frac{y_1 -
	y_2}{r^2}, \\
\dot{\eta}_j & = & v_y(x_j, y_j) + \frac{\kappa}{2\pi} \frac{x_1 -
	x_2}{r^2},
\end{eqnarray}
\end{subequations}
where $r^2 = (x_1 - x_2)^2 + (y_1 - y_2)^2 = (\xi_1 - \xi_2)^2 + (l +
\eta_1 - \eta_2)^2$.
The first terms on the right-hand side of Eq.~(\ref{eom}) are the velocity
field without the point vortices and the second terms are the velocity
field produced by the other point vortex.
We impose the same boundary condition as in Eq.~(\ref{negpot}) at $x^2 +
(y \pm D / 2)^2 = R^2$, giving the velocity field without the point
vortices as
\begin{equation} \label{vf2}
v_x - i v_y = \frac{4 U \sinh^2 \beta}{(z / R)^2 + \sinh^2 \beta} \left[
\wp(i \tau - i \beta) - \wp(i \tau + i \beta) \right],
\end{equation}
where $\beta$ and $\tau$ are defined in Sec.~\ref{s:double1}.
Substituting Eq.~(\ref{vf2}) into Eq.~(\ref{eom}) and neglecting the second
and higher orders of the small deviations, we have
\begin{subequations} \label{eom2}
\begin{eqnarray}
\dot{\xi}_1 & = & v_0 - c \eta_1 - \frac{\kappa}{2\pi} \left( \frac{1}{l}
	- \frac{\eta_1 - \eta_2}{l^2} \right), \\
\dot{\eta}_1 & = & -c \xi_1 + \frac{\kappa}{2\pi}
\frac{\xi_1 - \xi_2}{l^2}, \\
\dot{\xi}_2 & = & v_0 + c \eta_2 - \frac{\kappa}{2\pi} \left( \frac{1}{l}
	- \frac{\eta_1 - \eta_2}{l^2} \right), \\
\dot{\eta}_2 & = & c \xi_2 + \frac{\kappa}{2\pi}
\frac{\xi_1 - \xi_2}{l^2},
\end{eqnarray}
\end{subequations}
where $v_0 = v_x(0, \pm l / 2)$ and $c = -\partial_y v_x(0, l / 2) =
\partial_y v_x(0, -l / 2) = -\partial_x v_y(0, l / 2) = \partial_x v_y(0,
-l / 2)$.
In order for the point vortices to be stationary at $(0, \pm l / 2)$, we
set $v_0 = \kappa / (2 \pi l)$, which determines $l$ for given $U$, $D$,
and $R$.
Equations~(\ref{eom2}) then become homogeneous differential equations
and have the solutions of the form $\xi_j, \eta_j \propto e^{\lambda t}$.
The eigenvalues $\lambda$ are obtained as
\begin{equation} \label{lam}
\lambda = \pm \sqrt{c^2 - \frac{c \kappa}{\pi l^2}}.
\end{equation}
If the eigenvalues are pure imaginary, i.e., $c < \kappa / (\pi l^2)$ the
solutions are oscillatory and the two point vortices are stable.
If $c > \kappa / (\pi l^2)$, the small deviations grow exponentially and
the system is dynamically unstable.

\begin{figure}[t]
\includegraphics[width=8.5cm]{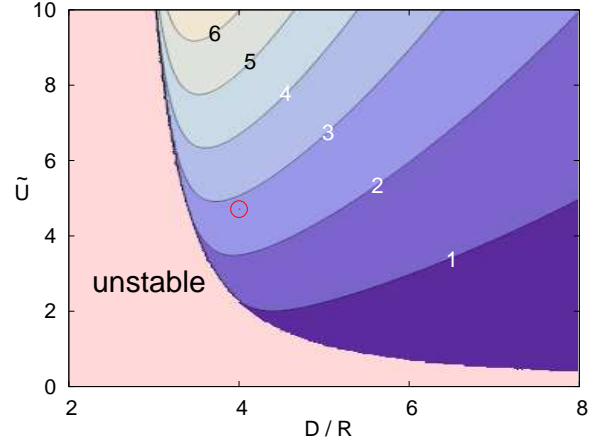}
\caption{
Stability of a trapped vortex dipole with respect to the
distance $D$ between the potentials and the normalized velocity $\tilde
U = 2 \pi R U / \kappa$.
When $\lambda$ in Eq.~(\ref{lam}) is real, the system is dynamically
unstable.
The contour plot shows $2 \pi R ^2 |\lambda| / \kappa$ where $\lambda$ is
pure imaginary.
The circle corresponds to the case in Fig.~\ref{f:trap}.
}
\label{f:stability}
\end{figure}
Figure~\ref{f:stability} shows a contour plot of $|\lambda|$ in
Eq.~(\ref{lam}) for the stable region.
The stable trapping dynamics shown in Fig.~\ref{f:trap} corresponds to the
circle in Fig.~\ref{f:stability}.
From Fig.~\ref{f:stability}, we find that $|\lambda|$ increases with an
increase in $U$ and the trapping is tight for large $U$.
However, if $U$ exceeds the critical velocity, vortex dipoles are
successively generated from the potentials as in Fig.~\ref{f:single},
which disturbs the trapping.
Therefore, there is an appropriate region of $D$ and $U$ for the most
efficient trapping.

\section{Manipulation of vortex dipoles}
\label{s:manip}

\begin{figure}[t]
\includegraphics[width=8.5cm]{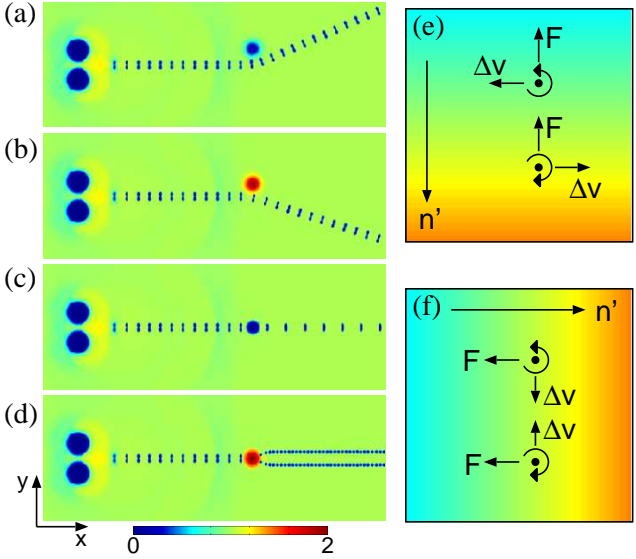}
\caption{
Stroboscopic images of the density profile $|\psi|^2 / n_0$ in the rest
frame.
A vortex dipole is launched by the two Gaussian potentials as in 
Fig.~\ref{f:launch2} with $V_0 / (g n_0) = 5$, $d / \xi = 4.5$, and $D /
\xi = 19$.
The stroboscopic images are taken at time intervals of $\Delta t / \tau =
20$.
(a) A Gaussian potential with $V_0 / (g n_0) = 1$ and $d / \xi = 4.5$ is
located at $y = 1.6$. 
(b) A Gaussian potential with $V_0 / (g n_0) = -1$ and $d / \xi = 4.5$ is 
located at $y = 1.6$. 
(c) A Gaussian potential with $V_0 / (g n_0) = 1$ and $d / \xi = 4.5$
is switched on at $t / \tau = 330$.
(d) A Gaussian potential with $V_0 / (g n_0) = -1$ and $d / \xi = 4.5$
is switched on at $t / \tau = 330$.
The field of view of each panel is $224 \xi \times 80 \xi$.
(e), (f) Schematic illustrations for explaining the Magnus effect.
}
\label{f:manip}
\end{figure}
We next show that the velocity and trajectory of a launched vortex dipole
can be changed after the launch.
We use a static Gaussian potential to curve trajectories of vortex
dipoles, and use a time-dependent Gaussian potential to accelerate and
decelerate them.
Figure~\ref{f:manip} shows stroboscopic images of such dynamics.
A vortex dipole is launched in the same manner as in Fig.~\ref{f:launch2}.
In Fig.~\ref{f:manip} (a), a weak repulsive Gaussian potential is located
near the trajectory of the vortex dipole and the trajectory bends toward
the potential.
In Fig.~\ref{f:manip} (b), the obstacle potential is attractive and the
trajectory bends in the opposite direction.
In Fig.~\ref{f:manip} (c), a repulsive Gaussian potential is switched on
at the moment that the vortex dipole passes through, which accelerates the
vortex dipole.
In Fig.~\ref{f:manip} (d), on the other hand, an attractive potential is
switched on and the vortex dipole is decelerated.

The behaviors in Figs.~\ref{f:manip} (a)-\ref{f:manip} (d) can be
understood by the Magnus effect.
The Magnus force on each vortex is given by $\bm{F}_{\rm M} = m n
\bm{\kappa} \times \Delta \bm{v}$, where $\bm{\kappa}$ is the vector of
circulation and $\Delta \bm{v}$ is the velocity relative to the
unperturbed velocity of the vortex dipole.
When force $\bm{F}$ is exerted on a vortex, the velocity changes in such a
way that $\bm{F}$ balances with $\bm{F}_{\rm M}$, and hence
\begin{equation}
\Delta \bm{v} = \frac{1}{m n \kappa^2} \bm{\kappa} \times \bm{F}.
\end{equation}
In Fig.~\ref{f:manip}, the origin of the force is the density gradient
produced by the obstacle Gaussian potential.
The energy of a vortex is roughly given by $E \sim \pi \hbar^2 n / m$ and 
the force on a vortex is $\bm{F} \sim -\pi \hbar^2 \nabla n / m$, which is
in the direction opposite to the density gradient.
Two examples of the effect of density gradient on a vortex dipole are
shown in Figs.~\ref{f:manip} (e) and \ref{f:manip} (f).
When the density gradient is perpendicular to the traveling direction of a
vortex dipole [rightward for Fig.~\ref{f:manip} (e)], its trajectory is
curved toward the direction of $-\nabla n$, which explains the dynamics in
Figs.~\ref{f:manip} (a) and \ref{f:manip} (b).
When a vortex dipole propagates in the direction of $\nabla n$ as in
Fig.~\ref{f:manip} (f), the distance between the two vortices decreases
and the vortex dipole accelerates.
This corresponds to the situation in Fig.~\ref{f:manip} (c), where a
density dip suddenly appears just behind the vortex dipole.
The deceleration in Fig.~\ref{f:manip} (d) can also be understood in a
similar manner.

\begin{figure}[t]
\includegraphics[width=8.5cm]{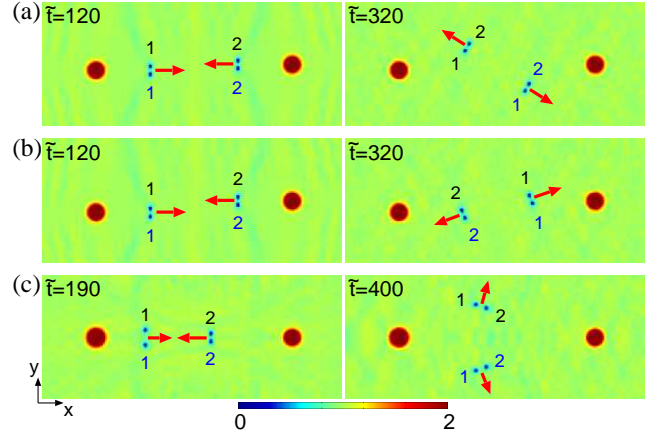}
\caption{
Collisions between vortex dipoles launched by the method in
Fig.~\ref{f:launch}.
The directions in which the vortex dipoles propagate are indicated by the
arrows.
The numbers and their colors identify the vortices.
(a) Collision with the impact parameter $b / \xi = 3.16$.
Vortex exchange occurs.
(b) Collision with $b / \xi = 6.32$.
The two vortex dipoles pass each other.
(c) Collision of vortex dipoles with different velocities.
The beam waist is $d / \xi = 3.54$ for the left-hand side in (c) and $d /
\xi = 3.16$ for the others.
The other parameters are the same as those in Fig.~\ref{f:launch}.
The field of view of each panel is $78 \xi \times 33 \xi$.
}
\label{f:collision}
\end{figure}
Using the method to launch vortex dipoles, we demonstrate collisions
between two counter-propagating vortex dipoles.
The two vortex dipoles are launched in the $\pm x$ directions by
red-detuned Gaussian beams as in Fig.~\ref{f:launch}.
Figures~\ref{f:collision} (a) and \ref{f:collision} (b) show collisions
between two vortex dipoles launched with the same velocity.
When the ``impact parameter'' $b$ is small, each vortex in the two vortex
dipoles are exchanged, and the two vortex dipoles with new pairs go away
with the same scattering angle [Fig.~\ref{f:collision} (a)].
The scattering angle is $\pi / 2$ for frontal collision ($b = 0$, data not
shown).
In Fig.~\ref{f:collision} (b), the impact parameter is large and the two
vortex dipoles pass each other without vortex exchange.
Figure~\ref{f:collision} (c) shows collision between vortex dipoles with
different velocities, where the vortex exchange occurs and the new vortex
dipoles scatter with scattering angles $\varphi$ and $\pi - \varphi$.
This is understood from the momentum conservation, where a slower vortex
dipole (larger distance between vortices) has a larger
momentum~\cite{Crescimanno}.
Collisions between point vortex dipoles in classical fluids have been
studied in Refs.~\cite{Manakov,Eckhardt}.

\section{Conclusions}
\label{s:conc}

We have investigated dynamics of a 2D BEC stirred by far-off-resonant
Gaussian laser beams.
When a red-detuned beam (attractive potential) moves through a BEC
exceeding a critical velocity, a vortex dipole is nucleated at the head of
the moving potential (Fig.~\ref{f:single}).
This is in contrast to a blue-detuned (repulsive) beam, in which a vortex
dipole is generated on both sides of the potential.
The vortex shedding from a red-detuned beam exhibits the hysteresis
behavior with respect to the velocity (Fig.~\ref{f:phase}).
When a red-detuned beam is displaced, a created vortex dipole propagates
towards the direction of the displacement, which can be used as a
``vortex-dipole launcher'' (Fig.~\ref{f:launch}).
When two repulsive potentials are moved, a vortex dipole is generated
between the potentials, which propagates to the direction opposite to the
motion of the potential (Fig.~\ref{f:double}).
A vortex dipole can also be launched by two blue-detuned beams
(Fig.~\ref{f:launch2}).
The velocity of vortex dipoles can be controlled in the two launching
methods.
For two red-detuned beams, we found that a vortex dipole can be trapped
between two moving potentials (Fig.~\ref{f:trap}).
Launched vortex dipoles can be manipulated by an external potential
(Fig.~\ref{f:manip}).
A trajectory of a vortex dipole is curved by a Gaussian potential located
near the trajectory.
A Gaussian potential with time-dependent intensity on the trajectory of a
vortex dipole can change its velocity.

We have thus shown that we can create and launch vortex dipoles in a
controlled manner and also manipulate their trajectories and velocities
after the launch.
These are achieved only by a quasi-2D BEC and Gaussian laser beams.
This technique enables us to investigate a variety of dynamics of
quantized vortices in superfluids, such as collisions of vortex dipoles
(Fig.~\ref{f:collision}).

\begin{acknowledgments}  
This work was supported by Grants-in-Aid for Scientific
Research (No.\ 20540388, No.\ 22340116, No.\ 23540464, and No.\ 23740307)
from the Ministry of Education, Culture, Sports, Science and
Technology of Japan, and Japan Society for the Promotion of
Science.
T. Kishimoto thanks for Special Coordination Funds for Promoting Science
and Technology (Highly Talented Young Researcher) from Japan Science and
Technology Agency.
\end{acknowledgments}

\end{document}